# Titanium diboride ceramics for solar thermal absorbers

**Elisa Sani\*, Marco Meucci, Luca Mercatelli**
*CNR-INO National Institute of Optics, Largo E. Fermi, 6, I-50125 Firenze, Italy*

**Andrea Balbo**
*Corrosion and Metallurgy Study Centre "Aldo Daccò", Engineering Department, University of Ferrara, G. Saragat 4a, Ferrara 44122, Italy & ISTEC-CNR, Institute of Science and Technology for Ceramics, Via Granarolo 64, Faenza 48018, Italy;*

**Clara Musa, Roberta Licheri, Roberto Orrù, Giacomo Cao**
*Dipartimento di Ingegneria Meccanica, Chimica e dei Materiali, Unità di Ricerca del Consorzio Interuniversitario Nazionale per la Scienza e Tecnologia dei Materiali (INSTM), Università degli Studi di Cagliari, Piazza D'Armi, 09123 Cagliari, Italy*

*\* Corresponding author, email:elisa.sani@ino.it*

## Abstract

Titanium diboride ($TiB_2$) is a low-density refractory material belonging to the family of ultra-high temperature ceramics (UHTCs). This paper reports on the production and microstructural and optical characterization of nearly fully dense $TiB_2$, with particular interest to its potential utilization as novel thermal solar absorber. Monolithic bulk samples are produced starting from elemental reactants by a two-step method consisting of the Self-propagating High-temperature Synthesis (SHS) followed by the Spark Plasma Sintering (SPS) of the resulting powders. The surface of obtained samples has-been characterized from the microstructural and topological points of view. The hemispherical reflectance spectrum has been measured from 0.3 to 15 μm wavelength, to evaluate the potential of this material as solar absorber for future concentrating solar plants.

1. Introduction

Solar thermal technology is recognized among the most promising renewable energy sources in the future. However, physics states that the efficiency of thermal cycles increases with temperature. For Concentrating Solar Power (CSP) plants, temperatures are usually limited to 800 K or lower values [1, 2], because of criticalities of the sunlight receiver element. For this reason, the main challenge for CSP technology advance is to find a receiver material able to withstand to operating temperatures higher than those allowed by current systems, while showing a high sunlight absorption as well as low re-radiation losses.

Recently, the use of carbide and boride materials belonging to the class of ultra-high temperature ceramics (UHTCs) was proposed for novel solar receivers operating at higher temperature than standard systems [3-11]. In fact, it was demonstrated that this material family shows intrinsic spectral selectivity which makes them appealing for sunlight absorption up to very high temperatures with reduced thermal losses [3-11]. Moreover, these materials also have well known characteristics such as ultra-refractoriness, good chemical stability at high temperature, good thermal conductivity, hardness and superior mechanical properties [12-15], which have implied their successful use for aerospace, military and, in general, for all applications where high temperatures conjugate extremely demanding performances [12- 18].

In this regard, titanium diboride ($TiB_2$) has been studied in the literature for its high hardness, low density, high melting point exceeding 3000°C, high wear resistance, high thermal and electrical conductivity and low thermal expansion coefficient [19, 20]. All these properties made this system attractive for ballistic armors, wear parts, cutting tools, etc. [21]. $TiB_2$ is also widely used in combination with other oxide and non-oxide ceramics, to increase strength and fracture toughness of the matrix [22-26].

Bulk monolithic $TiB_2$ materials are typically obtained by classical Hot-Pressing (HP) either starting from commercially available [27-34] or lab-made powders [32, 35-40].

In general, regardless the method used to synthesize the powders to be sintered, temperature levels equal or exceeding 1800°C and processing times on the order of hours are needed when using the HP approach to achieve densities levels above 95% of the theoretical value.

This holds also true when the $TiB_2$ powders to be hot-pressed were prepared by Self-propagating High-temperature Synthesis (SHS) [35]. Indeed, relative densities of 98% or more were reached only when operating at temperatures equal or higher than 1800°C. The only exception is represented by the study conducted by Peters et al. (2009) [32], where 98.6% and 98.9% dense products were obtained by HP at 1500°C (106 MPa, 1h) when starting from commercial $TiB_2$ powders mechanically treated for 30 min or using elemental reactants milled for 6h, respectively. However, iron contamination from milling media, i.e. 0.86 and 1.55 wt.%, respectively, was found in the milled powders.

Alternative sintering methods such as Hot Isostatic Pressing (HIP) [41], high-pressure sintering [42], high-pressure self-combustion synthesis [42], microwave sintering [43] and Spark Plasma Sintering (SPS) [44-50] have been also recently proposed for the fabrication of dense $TiB_2$ materials.

In this context the SPS technique, also referred to as Pulsed Electric Current Sintering (PECS), where the powders undergoing consolidation are rapidly heated by an electric pulsed current flowing through the conductive die and a mechanical load is simultaneously applied along the axial direction, was demonstrated to be particularly promising [51]. The various studies conducted so far clearly evidenced the capability of the latter technology to lead to highly dense $TiB_2$ products under relatively milder sintering conditions, with respect to the other consolidation methods previously mentioned.

Despite the high interest in $TiB_2$, its optical properties are, to the best of our knowledge, totally unexplored, as far as bulk materials are concerned. Indeed, the only literature source is limited to the spectral range from 0.4 to 1.0μm and is referred to thin films [52].

The present investigation is first aimed to the optimization of the SPS conditions for the full densification of additive free $TiB_2$ powders synthesized by SHS. In this regard, it should be noted that the combination of the SHS and the SPS techniques was recently exploited for the fabrication of other UHTC systems, both in monolithic [11, 53, 54] and composite forms [55-59].

Subsequently, in the present work we report on microstructure, topological characterization, and hemispherical reflectance spectra in the wavelength range 0.3-15 μm of $TiB_2$ produced by the two-steps SHS-SPS technique, with the aim to evaluate the material potential for solar absorber applications.

## 2. Experimental

Commercially available titanium (Sigma-Aldrich, St. Louis, Mo, USA, <45 μm, 99.98% purity), and amorphous boron (Sigma-Aldrich, St. Louis, Mo, USA, <1 μm, ≥ 95% purity) were used as starting powders for the synthesis of $TiB_2$ by SHS according to the following stoichiometry:

$$Ti + 2 B \rightarrow TiB_2 \qquad (1)$$

Reactants mixing was carried out in a SPEX 8000 (SPEX CertiPrep, USA) shaker mill for 20 min using plastic vials and six zirconia balls with 2 mm diameter. Approximately 8g of the obtained mixture was subsequently cold-pressed to obtain cylindrical pellets with a diameter of 10 mm and height of 30 mm, to be reacted by SHS under Ar atmosphere inside a closed stainless steel vessel. The reaction was activated at one pellet end using an electrically heated tungsten coil. A two-color pyrometer (Ircon Mirage OR 15-990, USA) was used for measuring the combustion temperature during SHS. The resulting porous material was reduced in powder form by milling about 4g of it for 20 min using the SPEX 8000 device with a ball-to-powder weight ratio of 2. A laser light scattering analyser (CILAS 1180, France) was utilized to determine particle size of obtained powders. Surface area was obtained through BET measurements performed using a Micromeritics ASAP 2020 equipment (Micromeritics, Georgia, USA).

Consolidation of SHS powders to produce $TiB_2$ cylindrical disks (about 14.7 mm diameter, and 3mm thickness) for optical characterization was carried out by Spark Plasma Sintering (SPS 515S model, Fuji Electronic Industrial Co., Ltd., Kanagawa, Japan). This apparatus consists of a uniaxial press, able to provide up to 50 kN loads, combined with a DC pulsed current generator (10 V, 1500 A, 300 Hz). A sequence of 12 ON pulses followed by 2 OFF pulses is adopted, with the characteristic time of single pulse equal to about 3.3 ms.

About 3 g of powder mixture were placed inside the graphite mould (outside diameter, 30 mm; inside diameter, 15 mm; height, 30 mm). Commercial $TiB_2$ powders (Sigma-Aldrich, St. Louis, Mo, USA, cod. 33628-9, < 10 μm) were also processed by SPS for the sake of comparison. A graphite foil (99.8 % pure, 0.13 mm thick, Alfa Aesar, Karlsruhe, Germany) was inserted between the internal surfaces of the die and the top and bottom surface of the sample and the plungers, to facilitate sample release at the end of the SPS process. Both die and plungers were made of AT101 graphite (Atal Srl., Italy). In addition, with the aim of minimizing heat losses by thermal radiation, the die was covered with a layer of graphite felt. The die was then placed inside the reaction chamber of the SPS apparatus and the system was evacuated down to 10-20 Pa.

During SPS experiments, the current was increased from zero at a constant rate up to a maximum intensity value ($I_M$) in 10 min. The latter level was maintained for a given holding time ($t_D$). The effects of $I_M$ and $t_D$ on powders densification were investigated in the ranges 800-1100 A and 0-20 min, respectively. The mechanical pressure (P) was kept constant to 60 MPa during the entire sintering process. The temperature of the external surface of the graphite mould was measured by an infrared pyrometer (Ircon Mirage OR 15-990, USA) focused on the lateral surface of the die. Each SPS run was repeated at least twice.

The relative density of the polished sintered samples was determined by the Archimede's method, using high purity distilled water as buoyant, at 20°C. Weighting of the specimen was carried out taking advantage of a Ohaus Explorer Pro (Ohaus Corporation, NJ, USA) analytical balance (± 0.0005 g precision), using the theoretical value of 4.5 g/cm$^3$ as reference for $TiB_2$ [19].

Phase identification was performed using a X-rays diffractometer (Philips PW 1830, Almelo, The Netherlands)with Cu Kα radiation (λ=1.5405 Å) and a Ni filter. A Rietveld analytical procedure was employed to estimate the relative amount of the phases present in SHS-obtained powders [60].

The microstructure of end products was examined by High-Resolution Scanning Electron Microscopy (HRSEM, mod. S4000, Hitachi, Tokyo, Japan), coupled with energy dispersive X-rays spectroscopy (EDS) (Thermo Fisher Scientific, Waltham, MA, USA).A ZEISS EVO LS 15 apparatus (Carl Zeiss Microscopy GmbH, Jena, Germany) equipped with a $LaB_6$ filament as electron source was also used for microstructural characterization.

The surface texture characterization was carried out with a non-contact 3D profilometer (Taylor-Hobson CCI MP, Leicester, UK) equipped with a 20X magnification objective lens. For each samples, two distinct areas (0.08 x 1 $cm^2$) were scanned along two orthogonal directions and the obtained 3D data were processed with the software Talymap 6.2 (Taylor-Hobson, Leicester, UK). In this work, the texture characterization was performed in terms of areal field parameters, as 3D parameters can provide a more comprehensive information about surface texture with respect to 2Dones. Thus, the evaluation of 3D texture parameters [61] was carried out on the two datasets collected for each sample, after denoising (median filter 5x5), form removing, S-filtering and after applying an areal robust gaussian L-filter (L-filter=0.8 mm).

Hemispherical reflectance was measured using a double-beam spectrophotometer (Perkin Elmer Lambda900) equipped with a Spectralon®-coated integration sphere for the 0.25-2.5 μm wavelength region and a Fourier Transform spectrophotometer (Bio-Rad "Excalibur") provided with a gold-coated integrating sphere and a liquid nitrogen-cooled detector for the range 2.5-15.0μm.

## 3. Results and discussion

### 3.1. Powders synthesis and characterization

A recent study addressed to the formation and simultaneous consolidation of $ZrB_2$ by reactive SPS evidenced the possible problems arising when the synthesis reaction of strongly exothermic systems takes place under the combustion regime inside a closed graphite mould [62]. This feature holds also true when considering titanium diboride, which is characterized by a very high formation enthalpy, i.e. $(-\Delta H_f^0)$ = 323.8 kJ/mol [63]. To overcome such drawbacks, in the present work synthesis and consolidation of $TiB_2$are carried out in two separate steps, the first one consisting in obtaining the boride phase by SHS according to reaction (1).

Upon ignition, the generated combustion front exhibited a self-sustaining character with a measured maximum temperature of about 2200°C. As shown in **Figure 1a**, where the XRD pattern of the resulting product is reported, during SHS the elemental reactants are almost completely transformed into hexagonal $TiB_2$. Small amounts of cubic and orthorhombic TiB were also detected by the XRD analysis, whereas no residual titanium was found. More specifically, the Rietveld analysis provided that the relative amount of $TiB_2$ in the SHS-obtained product was above 96 wt.%. As recently reported in the literature to justify the presence of unreacted Hf and Zr, respectively, in $HfB_2$ [59] and $ZrB_2$ [58] ceramics obtained by SHS, the formation of TiB is likely due to some deficiency of B in the reaction environment. This could be caused by some expulsion of the latter reactant during the evolution of the SHS reaction and/or by their partial consumption to reduce some oxides often present on the starting powders surface. Irrespective of the specific motivation, the reactants proportion required to produce $TiB_2$ according to Eq. (1) is not fulfilled, so that the following reactions might also take place [64]:

$$Ti + B \rightarrow TiB \quad (2)$$
$$Ti + TiB_2 \rightarrow 2TiB \quad (3)$$

In this regard, it should be noted that the use of some extra boron, i.e the boron-to-metal B/Me (Me=Hf, Zr) atomic ratio in the range 2.1-2.2, was found beneficial to decrease the amounts of residual reactants and secondary phases in $HfB_2$ [59] and $ZrB_2$ [58] products. On the other hand, all the attempt made along the same direction did not determine an improvement of the composition of the $TiB_2$-based ceramic synthesized in this work. Nonetheless, the reasonably high purity level achieved with an initial B/Ti molar ratio equal to 2 was considered satisfactory for the scope of the present investigation.

The obtained SHS products were characterized, after being pulverized and before their consolidation, by laser light scattering analysis, SEM and BET measurements. Granulometry data indicated that particles size was less than 20 μm, with an average value of 7.56±0.05 μm. Such result is consistent with SEM observations (cf. **Figure 1b**), which confirmed that each individual particle is generally smaller than 10 μm and most of them are few microns in size. In addition, BET analysis provides a surface area of 1.28 $m^2/g$ for $TiB_2$ powders produced in the present work. The latter value is slightly higher compared to those of $ZrB_2$ and $TaB_2$ powders recently synthesized following the same processing route, i.e. 1.09 and 1.16 $m^2/g$, respectively [58].

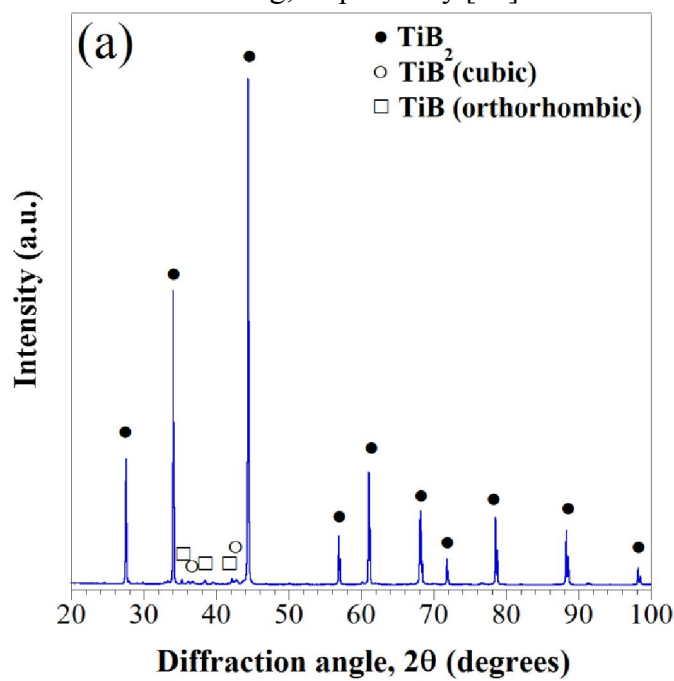

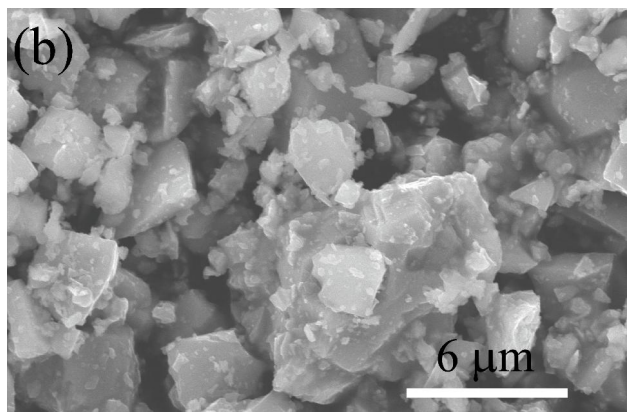

**Figure 1**. XRD pattern (a) and SEM micrograph (b) of TiB$_2$ powders obtained by SHS after 20 min ball milling.

*3.2. Sintering optimization and microstructure of dense samples*

The identification of the optimal SPS conditions for complete consolidation of TiB$_2$ powders produced by SHS was obtained investigating the effect of current intensity and holding time, while maintaining the applied pressure constant to 60 MPa. The obtained data are plotted in **Figure 2a** and **Figure 2b**, respectively.

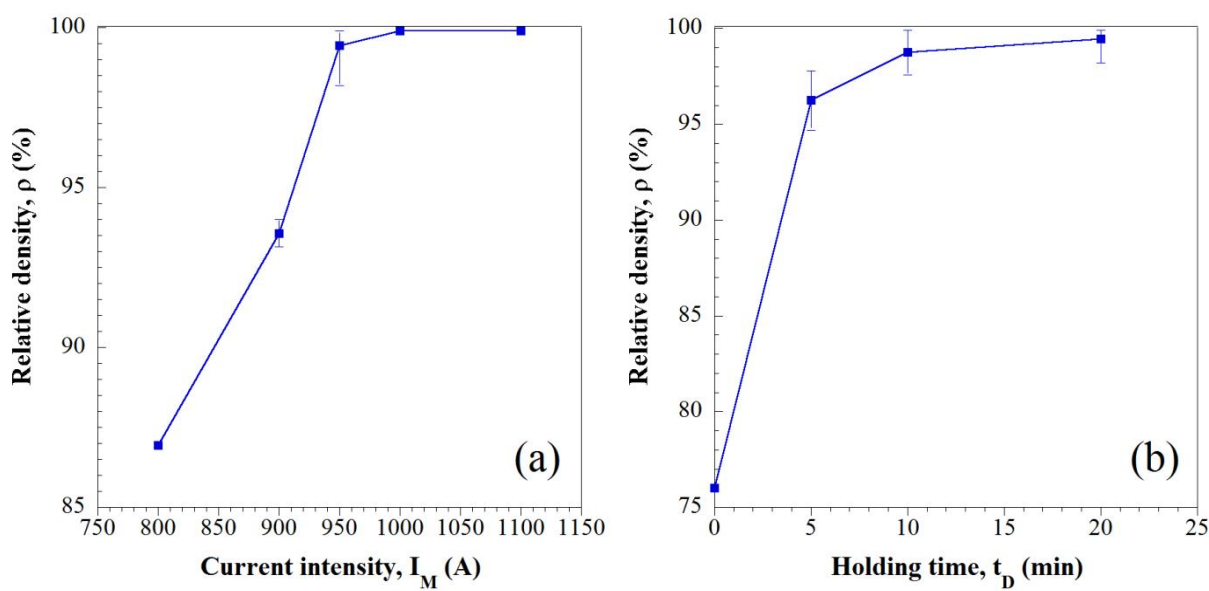

**Figure 2**. Effect of (a) mean current intensity ($t_D$= 20 min, P=60 MPa) and (b) dwell time ($I_M$= 950 A, P=60 MPa) on the density of TiB$_2$ bulk products obtained by SPS.

As shown in **Figure 2a**, an increase of the current intensity in the range of 800-950 A, which corresponds to a temperature rise approximately from 1400 to 1530°C, produces a significant improvement in powder densification. In particular, sintered products with relative densities of 99.45% (average value) were obtained when $I_M$=950 A was applied for 20 min. A further increase of the applied current to 1000 A or higher levels (T≥1575°C) leads to full consolidation.

As far as the study of the influence of holding time is concerned, the obtained results (**Figure 2b**) evidenced that the major effect is shown when processing powders are maintained at the highest current level for the first 5 min, where the relative density correspondingly increased from 76% to above 95% of the theoretical value. This effect proceeds, but at a lower rate, as sintering time was prolonged to 20 min. Based on the obtained results, the selected SPS conditions to produce nearly fully dense $TiB_2$ samples for optical measurements are $I_M$= 950 A, $t_D$= 20 min, and P=60 MPa. Correspondingly, the maximum temperature level measured during the sintering process was 1530±20°C. For the sake of comparison, the latter condition was also applied to process by SPS commercially available $TiB_2$ powders. The resulting sintered samples were highly porous, i.e. relative density of 80±3%, which is consistent with similar results reported in the literature for such system when operating with sintering temperature of about 1500°C [44]. It should be noted that the condition adopted in this work is relatively milder, not only with respect to those required when considering the classical HP approach, as mentioned in the Introduction, but also when the comparison is extended to previous studies involving SPS-like apparatuses for the fabrication of dense titanium diboride. For instance, 97.6% dense samples were obtained from commercial $TiB_2$ powders sintered by SPS at 1800°C for 5 min [44]. Even lower density levels, i.e. slightly higher than 80% [47] and 96% [50], were achieved at 1800°C and 2000°C, respectively, in similar studies. The characteristics of the starting powders apparently affect in a significant manner their sintering ability as well as the characteristics of resulting bulk product. A peculiar behavior was recently observed during the SPS process of $TiB_2$ powders previously synthesized by borothermal reduction of $TiO_2$ [46]. Indeed, in the cited reference it was unexpectedly observed that the relative density of sintered materials was lowered from 97.8 to 96.1% as the sintering temperature was increased from 1400 to 1500°C. It was postulated that this feature could be associated to the formation of a second phase (TiB) during SPS, which consumes $TiB_2$ thus leading to void formation. In contrast, as clearly shown in **Figure 2a**, an increase of the applied current above 950 A does not correspond in the present study to a decrease of the product density, as observed by Mukhopadhyay et al. [46].

As far as favourable densification conditions required in the present work are concerned, they can be ascribed not only to the use of the efficient SPS technique but also to the characteristics of SHS powders, which apparently display a high sintering ability. The latter characteristic was clearly evidenced for the $TiB_2$ system by Khanra et al. (2005) [65], who compared the sintering behavior displayed by SHS-treated powders obtained from $TiO_2$, boric acid and Mg, with respect to that of commercially available products. In particular, about 97 and 86% dense materials were respectively obtained when the two kind of powders were processed for 30 min at 1950°C in a furnace. This outcome is consistent with the results more recently reported in other studies involving the consolidation by SPS of other borides-based UHTCs [56, 62]. Such characteristics in the powders can be associated to the high defect concentration generated by the severe heating and cooling rate conditions (up to on the order of $10^5$ K/min) encountered during the evolution of the SHS process [66]. In addition, it is also likely that the small amount of TiB present in the powders synthesized in this work (cf. **Figure 1a**) might also help to promote their consolidation.

The microstructures of polished and fractured surfaces of dense products were first examined by SEM and two representative micrographs are shown in **Figure 3(a)** and **3(b)**, respectively.

Surprisingly, in spite of the rather high relative density achieved, the material surface (**Figure 3(a)**) appears to contain a not negligible amount of residual porosity (black areas). Nonetheless, the fracture surface reported in **Figure 3(b)** provides a completely different picture, as the bulk of the sample is almost fully dense with only few isolated pores. The most likely explanation for justifying the presence of observed surface voids is that a certain amount of small $TiB_2$ grains was removed during the polishing step. The latter statement is in agreement with the motivation recently provided by Sabahi Namini et al. (2015) [67] to explain the presence of pores on the surface of highly dense SiC-reinforced $TiB_2$ samples produced by hot-pressing.

The XRD analysis performed on the sample surface shown in **Figure 3(c)** evidenced for the SPS-treated product a composition similar to that of initial SHS powders, with $TiB_2$ as a major phase and small amounts of orthogonal and cubic TiB as by-products. The latter outcome is also confirmed by EDS analysis. Regarding grains size, **Figure 3(a)-3(b)** indicate that they generally range from few microns up to 15 μm, at most.

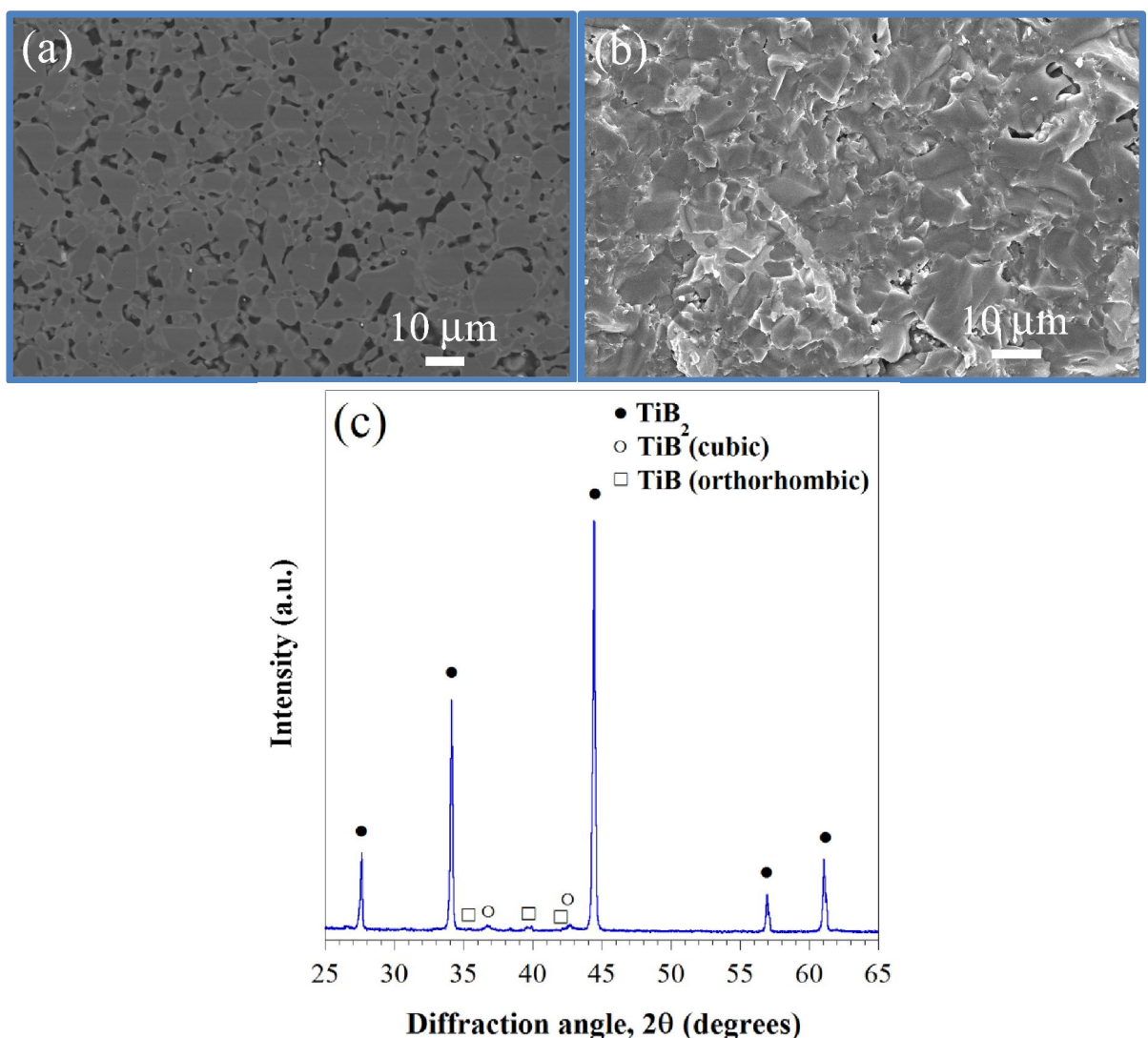

**Figure 3**. (a) Back-scattered SEM micrograph of the polished surface, (b) secondary electron image of the fracture surface, and (c) X-ray diffraction pattern of $TiB_2$ product obtained by SPS ($I_M$= 950 A, $t_D$= 20 min, P=60 MPa).

*3.3. Topological characterization*

The surface texture of a material has important effects on optical properties such as reflectance, absorbance and light scattering. Therefore, in order to investigate the intrinsic optical properties of the synthesized materials and evaluate the contributions arising from surface morphology, each sample was characterized from a topological point of view before performing optical measurements. The surface texture was characterized by evaluating the areal surface parameters on the SL surface (equivalent to the roughness surface) obtained by applying an L-filter on the SF surface.

The average values of the 3D surface texture parameter measured on the studied samples are reported in **Table 1**, along with the corresponding standard deviation. For a better evaluation of $TiB_2$ properties, two other previously investigated fully dense borides ($TaB_2$ and $ZrB_2$) [11] were considered as a term of comparison.

The $TaB_2$ sample showed values of Sa and Sq higher with respect to those measured on the other borides. Since these parameters are related to the surface roughness (Sa) and the way in which light is scattered from a surface (Sq), this result indicates a slightly higher surface energy of $TiB_2$ compared to $TiB_2$ and $ZrB_2$ samples [68].

All the borides are characterized by close values of both Ssk and Sku parameters. These parameters are related to the type of defects and to their distribution on the samples surface. In particular, negative values of Ssk indicate the predominance of pores or valleys structure, while large and positive values (>3) of Sku indicate the presence of a certain amount of high peaks or deep valleys/pores. These parameters provide useful information about the optical characteristics of a surface, in fact, negative values of Ssk and high values of Sku indicate a surface that can trap the incident photons and absorb them because of the multiple reflections from both sides of the pores/valleys.

The obtained results show that the studied borides have a quite similar surface texture, characterized by the prevalence of pronounced pores or valleys (Sv, maximum deep, in the range of 10-13 µm) and little amount of small peaks (Sp, maximum height, in the range of 1.4 – 3 µm).

**Table 1**: Measured average values and standard deviation of 3D surface texture parameters

| 3D parameters | Description | $TiB_2$ | $TaB_2$ | $ZrB_2$ |
|---|---|---|---|---|
| Sa (µm) | *Arithmetic mean height of the S-L Surface* | 0.237±0.057 | 0.440±0.042 | 0.211±0.028 |
| Sq (µm) | *Root mean square height of the S-L Surface* | 0.50±0.20 | 0.884±0.069 | 0.44±0.11 |
| Ssk | *Skewness of the S-L Surface* | -7.2±2.2 | -5.8±1.9 | -7.1±1.1 |
| Sku | *Kurtosis of the S-L Surface* | 94±40 | 51±29 | 83±15 |
| Sp (µm) | *Maximum peak height in the S-L Surface* | 1.44±0.47 | 1.46±0.27 | 2.9±2.0 |
| Sv (µm) | *Maximum pit height of the S-L Surface* | 10.6±3.7 | 12.9±3.0 | 9.6±2.6 |
| Sz (µm) | *Maximum height of the S-L Surface* | 12.0±4.1 | 14.3±2.8 | 12.5±4.6 |

*3.4. Optical properties*

**Figure 4** shows the acquired spectrum of $TiB_2$, together with those of $TaB_2$ and $ZrB_2$ for comparison [11]. The reflectance of $TiB_2$ nearly monotonically increases with wavelength, with values growing from 33% to 60% as the wavelength was increased in the range 0.3-2 µm. At about 5.5 µm, reflectance is about 86% and asymptotically approaches 100% towards the infrared.

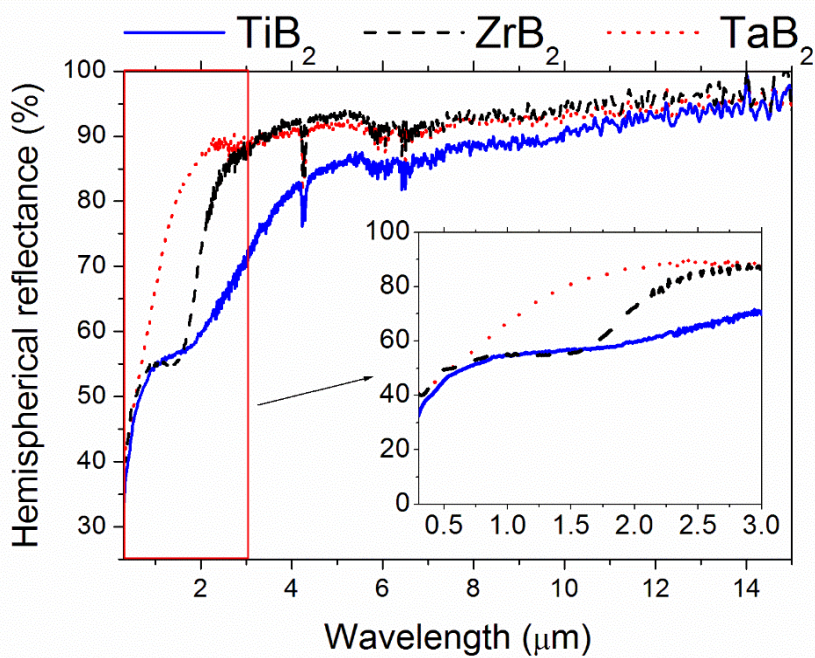

**Figure 4**: Experimental spectral hemispherical reflectance of $TiB_2$ (continuous blue line) compared to that of $ZrB_2$ and $TaB_2$.

When the comparison between the three borides is accounted for, we can notice that $TiB_2$ has a reflectance nearly superimposed to that of $ZrB_2$ and considerably lower than that of $TaB_2$ at wavelengths below 1.6 µm (see inset in **Figure 4**), a lower reflectance than other borides in the intermediate spectral region (1.6-6.0 µm) and a similar reflectance plateau at longer wavelengths. The spectral shapes and values of optical reflectance spectra are determined by different parameters: fundamental optical properties of the material (i.e. intrinsic properties connected to the chemical composition of the investigated surface) and morphological characteristics of the surface. From the morphological analysis of the surface reported in the previous paragraph, we can infer that the obtained spectral differences of the three borides directly arise from intrinsic characteristics of the different materials. In fact, the samples show very similar surface properties. In particular, they all are almost fully dense and have similar surface topologies. As for the roughness (Sa and Sq parameters), $TiB_2$ and $ZrB_2$ are very similar, while $TaB_2$ is characterized by a higher roughness. Fixed all other parameters, the effect of increasing roughness would be to decrease the optical reflectance. However, $TaB_2$ is the boride showing the highest reflectance for wavelengths shorter than 3 µm, while the small differences with respect to $ZrB_2$ at longer wavelengths lie within instrumental uncertainty and can be considered not significant. Thus we can conclude that the lowest reflectance of $TiB_2$ with respect to other borides, its smoother risefront towards infrared wavelengths and the slightly lower infrared reflectance plateau are intrinsic characteristics of this boride matrix. This fact affects sunlight absorption and thermal emittance properties of $TiB_2$, as discussed in the following.

From the acquired hemispherical reflectance spectrum $\rho^{\cap}(\lambda)$, it is possible to calculate the integrated solar absorbance $\alpha$ using the equation:

$$\alpha = \frac{\int_{\lambda_{min}}^{\lambda_{max}} (1-\rho^{\cap}(\lambda)) \cdot S(\lambda) d\lambda}{\int_{\lambda_{min}}^{\lambda_{max}} S(\lambda) d\lambda} \qquad (4)$$

where $S(\lambda)$ is the Sun emission spectrum [69] and the integration is carried out between $\lambda_{min} = 0.3$ µm and $\lambda_{max} = 3.0$ µm. The integrated thermal emittance $\varepsilon$ at the temperature T can be evaluated as:

$$\varepsilon = \frac{\int_{\lambda 1}^{\lambda 2} (1-\rho^{\cap}(\lambda)) \cdot B(\lambda,T) d\lambda}{\int_{\lambda 1}^{\lambda 2} B(\lambda,T) d\lambda} \qquad (5)$$

where $B(\lambda,T)$ is the blackbody spectral radiance at the temperature $T$ of interest and $\lambda 1 = 0.3$ µm and $\lambda 2 = 15.0$ µm.

By comparing the values for $\alpha$, we can assess that $TiB_2$ is a better sunlight absorber than other borides ($\alpha=0.49$ for $TiB_2$, 0.4 for $TaB_2$ and 0.47 for $ZrB_2$). The importance of a high solar absorbance in spectrally selective materials has been pointed out in the literature [70]. As far as the thermal emittance $\varepsilon$ is concerned, $TiB_2$ is also the most emissive in comparison to the other diboride systems, due to the relatively lower reflectance value at mid-infrared wavelengths. Thus, the obtained spectral selectivity $\alpha/\varepsilon$ is 2.2 at 1000K and 1.7 at 1400K. These values are lower than those of dense monolithic $ZrB_2$ and $TaB_2$ [11] (3.9 and 4.0 at 1000K and 2.6 and 3.3 at 1400K, respectively), but still higher with respect to those of SiC ($\alpha/\varepsilon=1.0$ at 1400K [3]), which is the material currently used in solar furnaces [71], and only slightly lower than other previously investigated $MoSi_2$-added dense UHTCs, e.g. $ZrB_2$ ($\alpha/\varepsilon=2.0$ at 1400K [3]) and $TaB_2$ ($\alpha/\varepsilon=2.1$ at 1300K [7]).

## 4. Conclusion

The SHS and SPS techniques were combined in this work to produce bulk dense $TiB_2$ samples. In particular, the optimal SPS conditions identified for obtaining about 99.5% dense products were 1530 ±20°C, 20 min and 60 MPa. It should be noted that the latter conditions are relatively milder with respect to those reported in the literature, when considering the consolidation of commercially available $TiB_2$ powders either by conventional HP or SPS. In particular, only 80% dense materials were obtained when commercial $TiB_2$ powders were processed under the same SPS conditions. This fact also evidences the beneficial effect produced by the use of the SHS technique for powder synthesis. The obtained materials display a quite uniform microstructure with few micron-sized grains.

Furthermore, in view of the possible utilization of this system as solar absorber, the optical reflectance of bulk $TiB_2$ in the wavelength range 0.3-15 μm is evaluated for the first time to the best of our knowledge. The reflectance curve appears as step-like shaped, similarly to $ZrB_2$ and $TaB_2$. In comparison to other transition metal borides, $TiB_2$ is characterized by similar reflectance values in the wavelength range below, roughly, 2μm and above 6 μm and a lower reflectance in the intermediate region. The solar absorbance α is higher than that of other borides (e.g. $ZrB_2$ and $TaB_2$), while the spectral selectivity α/ε is higher than that of SiC and lower than $ZrB_2$ and $TaB_2$. However, if the comparison between $TiB_2$ and other borides is concerned, it should be mentioned also that $TiB_2$ has a lower density than $ZrB_2$ (4.5 g/cm$^3$ versus 6.11 g/cm$^3$) and considerably lower than $TaB_2$ (12.18 g/cm$^3$). The reduced weight is a considerable advantage for the proposed application of bulk solar absorbers for solar tower plants, where a large receiver must be firmly sustained at a considerable height. Finally, it should be mentioned that $TiB_2$ is appealing also from point-of-view of raw material procurement and cost, because titanium is much cheaper and easier to obtain than zirconium and tantalum.

**Acknowledgements**
This activity has been carried out in the framework of the FIRB2012-SUPERSOLAR (Programma "Futuro in Ricerca", prot. RBFR12TIT1) project funded by the Italian Ministry of Education, University and Research. E. S. gratefully acknowledges the Italian bank foundation "Fondazione Ente Cassa di Risparmio di Firenze" within the framework of the "SOLE-NANO" project (pratica n. 2015.0861). Thanks are also due to Mr. Massimo D'Uva and Mr. Mauro Pucci (CNR-INO) for technical assistance.The authors also thank Dr. Carla Cannas (Dipartimento di ScienzeChimiche e Geologiche, University of Cagliari, Italy) for BET measurements, Dr. Paola Meloni (Dipartimento di Ingegneria Meccanica, Chimica e dei Materiali, University of Cagliari, Italy) for granting the use of the SEM apparatus (ZEISS EVO LS 15), and Dr. Sebastiano Garroni (University of Burgos, Spain) for performing Rietveld analysis on XRD spectra of SHS powders.

**References**

[1] Y. Tian, C.Y. Zhao, A review of solar collectors and thermal energy storage in solar thermal applications. *Applied Energy* 104 (2013) 538-553.
[2] C.E. Kennedy, Review of Mid- to High-temperature Solar Selective Absorber Materials, National Renewable Energy Laboratory, Technical report, 2002, pp.1-51.
[3] D. Sciti, L. Silvestroni, L. Mercatelli, J.-L. Sans, E. Sani, Suitability of ultra-refractory diboride ceramics as absorbers for solar energy applications. *Sol. Energy Mater. Sol. Cells* 109 (2013) 8-16
[4] E. Sani, L. Mercatelli, J.-L. Sans, L. Silvestroni, D. Sciti, Porous and dense hafnium and zirconium ultra-high temperature ceramics for solar receivers, *Optical Mater.* 36 (2013) 163-168
[5] E. Sani, L. Mercatelli, P. Sansoni, L. Silvestroni, D. Sciti, Spectrally selective ultra-high temperature ceramic absorbers for high-temperature solar plants. *J. Renew. Sust. Energy* 4 (2012) 033104
[6] E. Sani, L. Mercatelli, F. Francini, J.-L. Sans, D. Sciti, Ultra-refractory ceramics for high-temperature solar absorbers, *Scripta Mater.* 65 (2011) 775-778

[7]  D. Sciti, Silvestroni, L.; Sans, J.-L.; et al., Tantalum diboride-based ceramics for bulk solar absorbers. *Sol. Energy Mater. Sol. Cells* 130 (2014) 208-216

[8]  D. Sciti, L. Silvestroni, D.M. Trucchi, E. Cappelli, S. Orlando, E. Sani, Femtosecond laser treatments to tailor the optical properties of hafnium carbide for solar applications, *Sol. Energy Mater. Sol. Cells* 132 (2015) 460-466

[9]  E. Sani, L. Mercatelli, M. Meucci, A. Balbo, L. Silvestroni, D. Sciti, Compositional dependence of optical properties of zirconium, hafnium and tantalum carbides for solar absorber applications, *Solar Energy* 131 (2016) 199-207

[10] E. Sani, Mercatelli, L., Meucci, M., Silvestroni, L., Balbo, A., & Sciti, D.. Process and composition dependence of optical properties of zirconium, hafnium and tantalum borides for solar receiver applications. *Sol. Energy Mater. Sol. Cells* 155 (2016) 368-377.

[11] E. Sani, L. Mercatelli, M. Meucci, A. Balbo, C. Musa, R. Licheri, R. Orrù, G. Cao, Optical properties of dense zirconium and tantalum diborides for solar thermal absorbers, *Renewable Energy*, 91 340-346 (2016)

[12] T. Ogawa, K. Ikawa, Diffusion of metal fission products in $ZrC_{1.0}$, *J. Nucl Mater.* 105 (1982) 331-334

[13] H. J. Ryu, Y. W. Lee, S. I. Cha. S. H. Hong, Sintering behaviour and microstructures of carbides and nitrides for the inert matrix fuel by spark plasma sintering, *J. Nucl Mater.* 352 (2006) 341-348

[14] W.G. Fahrenholtz, G.E. Hilmas, I.G. Talmy, J.A. Zaykoski, Refractory diborides of zirconium and hafnium, *J. Am. Ceram. Soc.*, 90 (2007) 1347–1364

[15] J. Buesking, Spain, A. Cull, J. Routbort, E. Gutierrez-Mora, Design for ultra high temperature applications: the mechanical and thermal properties of HfB2, HfCx, HfNx and Hf(N), *J. Mater. Sci.* 39 (2004) 5939–5949

[16] R. Loehman, E. Corral, H.P. Dumm, P. Kotula, R. Tandon, "*Ultra High Temperature Ceramics for Hypersonic Vehicle Applications*". Edited by Sandia National Laboratories Albuquerque, New Mexico and Livermore, California, report no. SAND2006-2925 (2006)

[17] A. Paul, J. Binner, B. Vaidhyanathan, UHTC Composites for Hypersonic Applications, in: Ultra-high temperature ceramics: materials for extreme environment applications, edited by William G. Fahrenholtz, Eric. J. Wuchina, William E. Lee, Yanchun Zhou (Wiley 2014)

[18] W. E. Lee, E. Giorgi, R. Harrison, A. Maître, O. Rapaud, Nuclear Applications forUltra-High Temperature Ceramics and MAX Phases, in: Ultra-high temperature ceramics : materials for extreme environment applications, edited by William G. Fahrenholtz, Eric. J. Wuchina, William E. Lee, Yanchun Zhou (Wiley 2014)

[19] J. F. Shackelford, W. Alexander, Materials Science and Engineering Handbook, CRC Press 2000

[20] R. G. Munro, Material Properties of Titanium Diboride, J. Res. Natl. Inst. Stand. Technol. 105 (2000) 709-720

[21] V. I. Matkovich, Boron and Refractory Borides. Springer, Berlin, 1977

[22] I. Bogomol, H. Borodianska, T. Zhao, T. Nishimura, Y. Sakka, P. Loboda, O. Vasylkiv, A dense and tough ($B_4C$–$TiB_2$)–$B_4C$ 'composite within a composite' produced by spark plasma sintering, *Scripta Mater.* Volume 71, 15 January 2014, Pages 17-20

[23] J. Oñoro, High-temperature mechanical properties of aluminium alloys reinforced with titanium diboride ($TiB_2$) particles, *Rare Metals* 30 (2011) 200-205

[24] Y. Ohya, M. J. Hoffmann, G. Petzow, Mechanical properties of hot-pressed SiC-$TiB_2$/TiC composites synthesized in situ, *J. Mat. Sci. Lett.* 12 (1993) 149-152

[25] S. Ran, O. Van der Biest, J. Vleugels, In situ platelet-toughened $TiB_2$–SiC composites prepared by reactive pulsed electric current sintering, *Scripta Mater.* 64 (2011) 1145-1148

[26] C. Musa, A. M. Locci, R. Licheri, R. Orrù, G. Cao, D. Vallauri, F. A. Deorsola, E. Tresso, J. Suffner, H. Hahn, P. Klimczyk, L. Jaworska, Spark plasma sintering of self-propagating high-temperature synthesized $TiC_{0.7}/TiB_2$ powders and detailed characterization of dense product, *Ceram. Int.* 35 (2009) 2587-2599

[27] J. Ho Park, Koh Y., Kim H., Hwang C., Kong E., ''Densification and Mechanical Properties of Titanium Diboride with Silicon Nitride as Sintering Aid,'' *J. Am. Ceram. Soc.* 82 [11] (1999), 3037–42;

[28] B. Basu, Vleugels J., Van der Biest O., Fretting wear behavior of $TiB_2$-based materials against bearing steel under water and oil lubrication, *Wear* 250 (2001) 631-641;

[29] R. Konigshofer, Furnsinn S., Steinkellner P., Lengauer W., Haas R., Rabitsch K., Scheerer M., Solid-state properties of hot-pressed $TiB_2$ ceramics, Int. *J. Refract. Met. Hard Mater.* 23 (2005), 350-357;

[30] L.H. Li, Kim H.E., Kang E.S., "Sintering and mechanical properties of titanium diboride with aluminum nitride as a sintering aid", *J. Europ. Ceram. Soc.* 22 (2002) 973-977;

[31] G. B. Raju, Basu B., "Densification, Sintering Reactions and properties of Titanium Diboride with titanium disilicide as a sintering aid", *J. Am. Ceram. Soc.* 90 [11] (2007) 3415-3423;

[32] J.S. Peters, Cook B.A., Harringa J.L., Russel A.M., Microstructure and low resistance of low temperature hot pressed $TiB_2$, Wear 266 (2009), 1171-1177;

[33] G.B. Raju, Basu B., Tak N.H., Cho S.J., Temperature dependent hardness and strength properties of $TiB_2$ with $TiSi_2$ sinter aid, *J. Eur. Ceram. Soc.* 29 (2009) 2119-2128;

[34] D. King, Fahrenholtz W., Hilmas G., Silicon Carbide-titanium diboride ceramic composites, *J. Europ. Ceram. Soc.* 33 (2013) 2943-2951

[35] W. Wang, Fu Z., Wang H., Yuan R., Influence of hot pressing sintering temperature and time on microstructure and mechanical properties of $TiB_2$ ceramics, *J. Eur. Ceram. Soc.* 22 (2002), 1045-1049;

[36] T. Murthy, Basu B., Balasubramaniam R., Suri A., Subramanian C., Fotedar R., Processing and properties of $TiB_2$ with $MoSi_2$ sinter-additive: a first report, *J. Am. Ceram. Soc.* 89 1 (2006), 131-138;

[37] T. Murthy, Subramanian C, Fotedar R., Gonal M., Sengupta P., Kumar S., Suri A., Preparation and property evaluation of $TiB_2$ + $TiSi_2$ composite, *Int. J. Refract. Met. Hard Mater.* 27 (2009) 629-636;

[38] A. Mukhopadhyay, Raju G., Basu B., Suri A., Correlation between phase evolution, mechanical properties and instrumented indentation response of $TiB_2$–based ceramics, *J. Eur. Ceram. Soc.*, 29 (2009) 505-516;

[39] A. Rabiezadeh A., Hadian A.M., Ataie A., Synthesis and sintering of $TiB_2$ nanoparticles, Ceramics International 40 (2014) 15775-15782;

[40] T.S.R.Ch. Murthy, Sonber J.K., Vishwanadh B., Nagaraj A., Sairam K., Bedse R.D., Chakravartty J.K., Densification, characterization and oxidation studies of novel $TiB_2$-$EuB_6$ compounds, *J. Alloys Compd.* 670 (2016) 85-95.

[41] Y. Muraoka, Yoshinaka M., Hirota K., Yamaguchi O., Hot Isostatic Pressing of TiB2-ZrO2 (2 Mol% Y2O3) composite powders, *Materials Research Bulletin* 31[7] (1996), 787-792

[42] S.K. Bhaumik, Divakar C., Singh A.K., Upadhyaya G.S., Synthesis and sintering of TiB2 and TiB2–TiC composite under high pressure, *Mater. Sci. Eng. A* 279 (2000) 275-281

[43] D. Demirskyi, Agrawal D., Ragulya A. Tough ceramics by microwave sintering of nanocristalline titanium diboride ceramics, *Ceram. Int.* 40 (2014) 1303-1310

[44] Z. Zhang, Shen X, Wang F, Lee S, Wang L., Densification behavior and mechanical properties of spark plasma sintered monolithic $TiB_2$ ceramics, *Mater. Sci. Eng. A* 527 (2010) 5947-5951;

[45] S. Ran, Zhang L., Van der Biest O., Vleugels J. Pulsed electric current, in situ synthesis and sintering of textured $TiB_2$ ceramics. *J. Europ. Ceram. Soc.* 30 (2010) 1043-1047

[46] A. Mukhopadhyay, Venkateswaran T., Basu B. Spark plasma sintering may lead to phase instability and inferior mechanical properties: A case study with $TiB_2$. *Scripta Mater.* 69 (2013) 159-164;

[47] A. Turan, Sahin F.C., Goller G., Yucel O., "Spark Plasma Sintering of monolithic $TiB_2$ ceramics", *Ceramic Processing Research* 15[6] (2014) 464-468;

[48] N.S. Karthiselva, Murty B.S., Srinivasa R. Bakshi, Low temperature synthesis of dense $TiB_2$ compacts by reaction spark plasma sintering, *Int. J. Refract. Met. Hard Mater.* 48 (2015) 201-210

[49] D. Demirskyi, Sakka Y, Vasylkiv O., High temperature reactive spark plasma consolidation of $TiB_2$-NbC ceramic composites, *Ceram. Int.* 41 (2015) 10828-10834;

[50] D. Demirskyi, Sakka Y, High temperature reaction consolidation of TaC-$TiB_2$ ceramic composites by spark plasma sintering, *J. Eur. Ceram. Soc.* 35 (2015) 405-410

[51] R. Orrù, R. Licheri, A.M. Locci, A. Cincotti, G. Cao, Consolidation/synthesis of materials by electric current activated/assisted sintering, Mater. Sci. Eng. R 63 (2009) 127-287.

[52] B.N. Beckloff, W. J. Lackey, Process–Structure–Reflectance Correlations for TiB2 Films Prepared by Chemical Vapor Deposition, *J. Am. Ceram. Soc.* 82[3] (1999) 503–512

[53] C. Musa C., R. Orrù, D. Sciti, L. Silvestroni, G. Cao, Synthesis, consolidation and characterization of monolithic and SiC whiskers reinforced $HfB_2$ ceramics, *J. Eur. Ceram. Soc.* 33 (2013) 603-614

[54] R. Licheri, C. Musa, R. Orrù, G. Cao, D. Sciti, and L. Silvestroni, Bulk Monolithic Zirconium and Tantalum Diborides by Reactive and Non-reactive Spark Plasma Sintering, *J. Alloys Compd.* 663 (2016) 351-359

[55] R. Licheri, R. Orrù, A. M. Locci, G. Cao, Efficient Synthesis/Sintering Routes to obtain Fully Dense $ZrB_2$-SiC Ultra-High-Temperature Ceramics (UHTCs), *Ind. Eng. Chem. Res.* 46 (2007) 9087-9096

[56] R. Licheri, R. Orrù, C. Musa, G. Cao, Combination of SHS and SPS Techniques for Fabrication of Fully Dense $ZrB_2$-ZrC-SiC Composites, *Mater. Letters* 62, 432–435 (2008);

[57] R. Licheri, R. Orrù, C. Musa, A.M. Locci, G. Cao, Consolidation via Spark Plasma Sintering of $HfB_2$/SiC and $HfB_2$/HfC/SiC Composite Powders obtained by Self-propagating High-temperature Synthesis, *J. Alloys Compd.* 478 572–578 (2009);

[58] R. Licheri, R. Orrù, C. Musa, and G. Cao, Efficient technologies for the Fabrication of dense $TaB_2$-based Ultra High Temperature Ceramics, *ACS Appl. Mater. Interfaces* 2(8) (2010) 2206-2212

[59] C. Musa, R. Licheri, R. Orrù, G. Cao "Synthesis, Sintering and Oxidative Behaviour of $HfB_2$-$HfSi_2$ ceramics" *Industrial & Engineering Chemistry Research* 53 (2014) 9101−9108

[60] L. Lutterotti, R. Ceccato, R. Dal Maschio, E. Pagani, Quantitative analysis of silicate glass in ceramic materials by the Rietveld method, *Mater. Sci. Forum*, 278-281 (1998) 87-92

[61] Geometrical Product Specifications (GPS)—Surface Texture: Areal—Part 2: Terms, Definitions and Surface Texture Parameters; ISO 25178-2:2012; ISO: Geneva, Switzerland; April 2012

[62] R. Licheri, C. Musa, R. Orrù, G. Cao, Influence of the heating rate on the in-situ synthesis and consolidation of $ZrB_2$ by Reactive Spark Plasma Sintering, *J. Eur. Ceram. Soc.* 35(4) (2015) 1129–1137

[63] I. Barin, Thermochemical data of pure substances. VHC, Weinheim, Germany, 1989.

[64] H. Feng, Zhou, Y., Jia, D., Meng, Q. Rapid synthesis of Ti alloy with B addition by spark plasma sintering. *Materials Science and Engineering A* 390(1-2) (2005) 344-349

[65] A.K. Khanra, Godkhindi, M.M., Pathak, L.C. Comparative studies on sintering behavior of self-propagating high-temperature synthesized ultra-fine titanium diboride powder *J. Am. Ceram. Soc.* 88(6) (2005) 1619-1621

[66] S.K. Mishra, S. Das, L.C. Pathak, Defect structures in zirconium diboride powder prepared by self-propagating high-temperature synthesis, *Mater. Sci. Eng. A* 364 (2004) 249-255

[67] A. Sabahi Namini, Seyed Gogani, S.N., Shahedi Asl, M., Farhadi K., Ghassemi Kakroudi, M. Microstructural development and mechanical properties of hot pressed SiC reinforced $TiB_2$ based composite. *Int. J. Refract. Met. Hard Mater.* 51 (2015) 169-179

[68] R.K. Leach, Fundamentals principles of engineering nanometrology, 2nd ed.; Elsevier: Amsterdam, The Netherlands, 2010

[69] Standard Tables for Reference Solar Spectral Irradiances: Direct Normal and Hemispherical on 37° Tilted Surface, Active Standard ASTM G173. ASTM G173–03(2012)

[70] K. Burlafinger, A. Vetter, C.J. Brabec, Maximizing concentrated solar power (CSP) plant overall efficiencies by using spectral selective absorbers at optimal operation temperatures, *Solar Energy* 120 (2015), 428–438

[71] C.C. Agrafiotis, I. Mavroidis, A.G. Kostandopoulos, B. Hoffschmidt, P. Stobbe, M Fernandez, V.Quero, Evaluation of porous silicon carbide monolithic honeycombs as volumetric receivers/collectors of concentrated solar radiation, *Sol. Energy Mater. Sol. Cells* 91 (2007) 474–488.